\documentclass[prb,aps,showpacs,superscriptaddress,amsmath,amssymb,floatfix,twocolumn]{revtex4-1}
\usepackage{graphicx}
\usepackage{color}
\usepackage{young}
\usepackage[colorlinks,bookmarks=false,citecolor=blue,linkcolor=red,urlcolor=blue]{hyperref}
\usepackage{amsfonts}
\usepackage{times}
\usepackage{changes}
\usepackage{wasysym}

\begin{document}
\title{SU(6) Heisenberg model on the honeycomb lattice: competition between plaquette and chiral order}
\date{\today} 

\author{Pierre Nataf}
\affiliation{Institute of Theoretical Physics, Ecole Polytechnique F\'ed\'erale de Lausanne (EPFL), CH-1015 Lausanne, Switzerland}
\author{Mikl\'os Lajk\'o}
\affiliation{Institute for Solid State Physics, University of Tokyo, Kashiwa 277-8581, Japan}
\author{Philippe Corboz}
\affiliation{University of Amsterdam, Faculty of Science ITF, NL-1090 GL Amsterdam}
\author{Andreas M. L\"auchli}
\affiliation{Institut f\"ur Theoretische Physik, Universit\"at Innsbruck, A-6020 Innsbruck, Austria}
\author{Karlo Penc}
\affiliation{Institute for Solid State Physics and Optics, Wigner Research
Centre for Physics, Hungarian Academy of Sciences, H-1525 Budapest, P.O.B. 49, Hungary}
\affiliation{MTA-BME Lend\"ulet Magneto-optical Spectroscopy Research Group, 1111
Budapest, Hungary}
\author{Fr\'ed\'eric Mila}
\affiliation{Institute of Theoretical Physics, Ecole Polytechnique F\'ed\'erale de Lausanne (EPFL), CH-1015 Lausanne, Switzerland}

\begin{abstract}
We revisit the SU(6) Heisenberg model on the honeycomb lattice, which has been predicted to be a chiral spin liquid by mean-field theory [G. Szirmai {\it et al.}, Phys. Rev. A{\bf 84}, 011611 (2011)]. 
Using exact diagonalizations of finite clusters, infinite projected entangled pair states simulations, and variational Monte Carlo
simulations based on Gutzwiller projected wave functions, we provide strong evidence in favour of the competing plaquette state, 
which was reported to be higher but close by in energy according to mean-field theory. This is further confirmed by the investigation 
of the model with a ring exchange term, which shows that there is a transition between the plaquette state and the chiral state at
a finite value of the ring exchange term. 
\end{abstract}

\pacs{67.85.-d, 71.10.Fd, 75.10.Jm, 02.70.-c}

\maketitle

With the recent progress towards achieving SU($N$) symmetry with ultra-cold fermionic atoms,\cite{WuPRL2003,gorshkov2010,Scazza2014,takahashi2012,Pagano2014,ZhangScience2014} the investigation of the effective SU($N$) Heisenberg model on various 1D and 2D lattices has become a very active field of research. Several remarkable ground state properties have been reported, including long-range color order,\cite{toth2010} algebraic correlations,\cite{Corboz12_su4} translational symmetry breaking valence-bond solid states in which groups of $N$ atoms form local singlets on plaquettes,\cite{CorbozSimplex2012,Corboz13_su3hc} and chiral ground states, suggested by Hermele {\it et al.}\cite{hermele2009,HermeleGurarie2011} for Mott insulators on square lattice with several particles per site. Interestingly, a mean-field calculation even predicted a chiral spin liquid in the SU(6) Heisenberg model on the honeycomb lattice with only one particle per site.\cite{SzirmaiG2011}
However, the rather natural plaquette state in which six SU(6) spins form singlets
on nonadjacent hexagons was found to lie very close in energy. 
So this result calls for further investigation with methods that go beyond mean-field theory.

In this paper, we have attacked this problem with state-of-the-art numerical methods: exact diagonalizations (ED), infinite projected entangled pair states simulations (iPEPS), and variational Monte Carlo (VMC) simulations based on Gutzwiller projected wave functions. As could be expected from the quasi-degeneracy of the mean-field results, it turned out to be very difficult to solve the problem, and all methods had to be pushed to their limit to reach a definitive conclusion, but each method led independently to the same conclusion that the ground state is actually a plaquette state. The chiral state is not far in parameter space however, and it does not take a large ring-exchange term to stabilize it, as demonstrated by ED and VMC.

The SU(6) Heisenberg model is defined by the Hamiltonian 
\begin{equation}
\mathcal{H}=\sum_{\langle i,j \rangle} P_{ij},
\label{eqn:HamP2}
\end{equation}
where the operator $P_{ij}= \sum_{\alpha,\beta}  |\alpha_i \beta_j \rangle \langle \beta_i \alpha_j |$ exchanges the $N=6$ colors $\alpha$ and $\beta$ of the atoms on neighboring sites $i,j$ of a honeycomb lattice.


\begin{figure}[b]
\includegraphics[width=0.99\columnwidth]{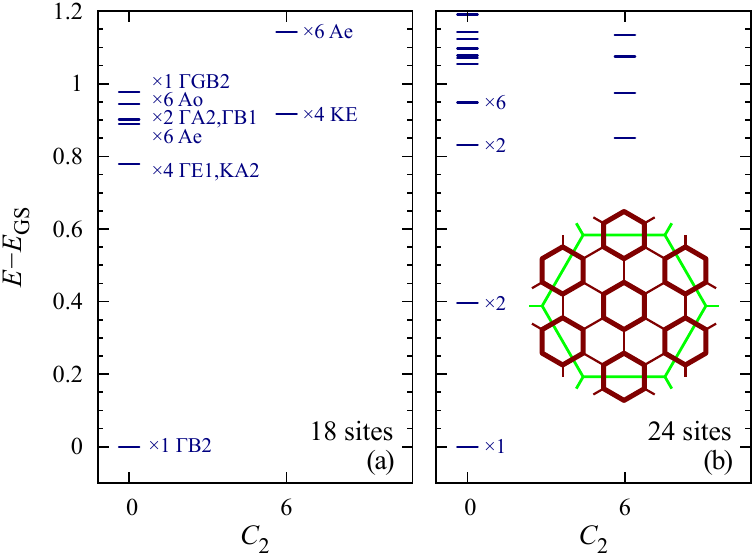}
\caption{Spectrum of the 18- (left) and 24-site (right) clusters as a function of the quadratic Casimir $\mathcal{C}_2$.
The degeneracies of some states are indicated, as well as the spatial quantum numbers for the 18-site cluster. 
For the 24-site cluster, the presence of 3 low-lying states is a strong indication of a plaquette phase (see text for details). 
Inset: broken-symmetry plaquette state reconstructed from ED. It breaks translations, but the $D_6$ symmetry is preserved. The bond energy is -0.81 (-0.56) for the thick (thin) lines.} 
\label{ED_24sites}
\end{figure}

\begin{figure*}[t]
\includegraphics[width=0.85\textwidth]{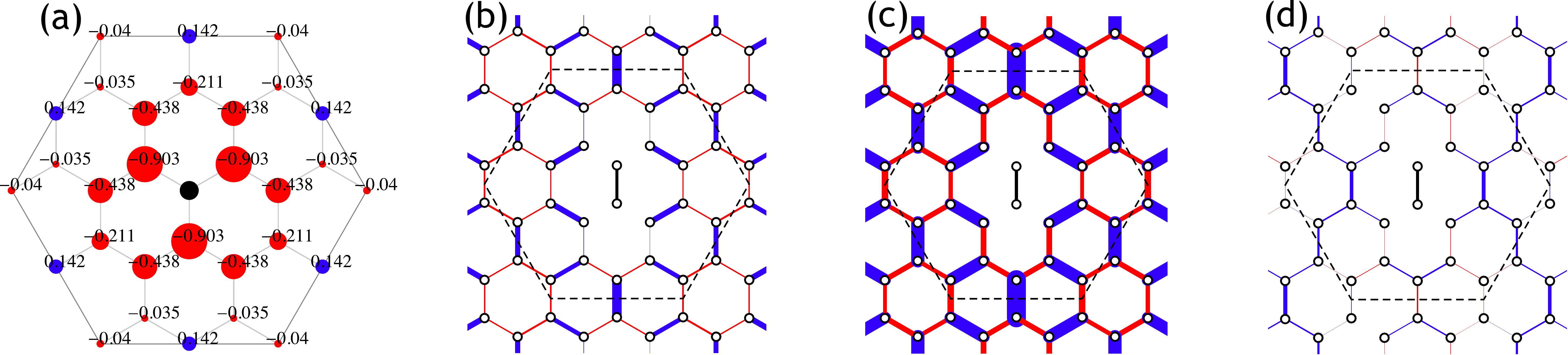}
\caption{$\left\langle P_{0i} \right\rangle - 1/6$ spin-spin (a) and  $\left\langle P_{12} P_{kl} \right\rangle - \left\langle P_{12}  \right\rangle \left\langle P_{kl}  \right\rangle$ dimer-dimer correlations (b)  in the exact ground-state of the 24-site cluster. As a reference we present the dimer-dimer correlations of  the translational invariant linear combination of  variational $0\pi\pi$-flux projected states with $\left|t_d/t_h \right|=0.8$ (c), and of the  variational $2\pi/3$-flux projected state (d). The pattern of the dimer-dimer correlations of the ED (b) and the $0\pi\pi$ variational states (c) is an indication of long-range plaquette ordering. 
}
\label{correlations_24sites}
\end{figure*}
{\it ED: }
With the standard exact diagonalization approach that takes into account all spatial symmetries but only an abelian subgroup of the SU($N$) symmetry group (color conservation plus cyclic color permutations), 
the currently largest accessible cluster with a number of sites multiple of 6 (a requirement for having a singlet ground state) is an 18-site cluster. The spectrum
is shown in Fig.~\ref{ED_24sites}(a). The plaquette state is expected to be 3-fold degenerate in the thermodynamic limit (one state at the $\Gamma$ point and two states at the two $K$ points in the Brillouin zone), but in the 18 site cluster the plaquettes can also wrap around the torus,\cite{Corboz13_su3hc} artificially enlarging the number of plaquette coverings to 6. By contrast , the chiral state is $2\times N=2\times 6=12$-fold degenerate in the spontaneous time-reversal symmetry (TRS) breaking scenario. While the first three levels $\Gamma B_2$, $KA_2(2\times)$ (plus the symmetry related level $\Gamma E_1$ particular to $N_s=18$) are in agreement with the expectations for a plaquette state,~\cite{Corboz13_su3hc} these
states are very close to many other excited states (including non singlets). So it is difficult to provide strong evidence for either 
state on the  basis of the 18-site cluster.

To go further, we have used a newly developed method that allows one to take advantage of the full SU($N$) symmetry, hence to work directly in the irreducible representations of SU($N$). For the singlet and the smallest values of the Casimir operator, this leads to Hilbert spaces of much smaller dimension than the standard approach.\cite{nataf2014} The spectrum is shown in Fig.~\ref{ED_24sites}(b). Interestingly enough, on 24 sites, the spectrum consists of 3 low-lying states reasonably well separated from the rest of the spectrum, the first indication that the ground state might have plaquette order. The spin-spin and dimer-dimer correlations are shown in  Fig.~\ref{correlations_24sites}. The spin-spin correlations decay quite fast, consistent with some kind of spin liquid, and the dimer-dimer correlations are consistent with a plaquette phase on the honeycomb lattice (see for instance the discussion of the SU(3) case in Ref.~[\onlinecite{Corboz13_su3hc}]). 

\begin{figure}[t]
\includegraphics[width=1\columnwidth]{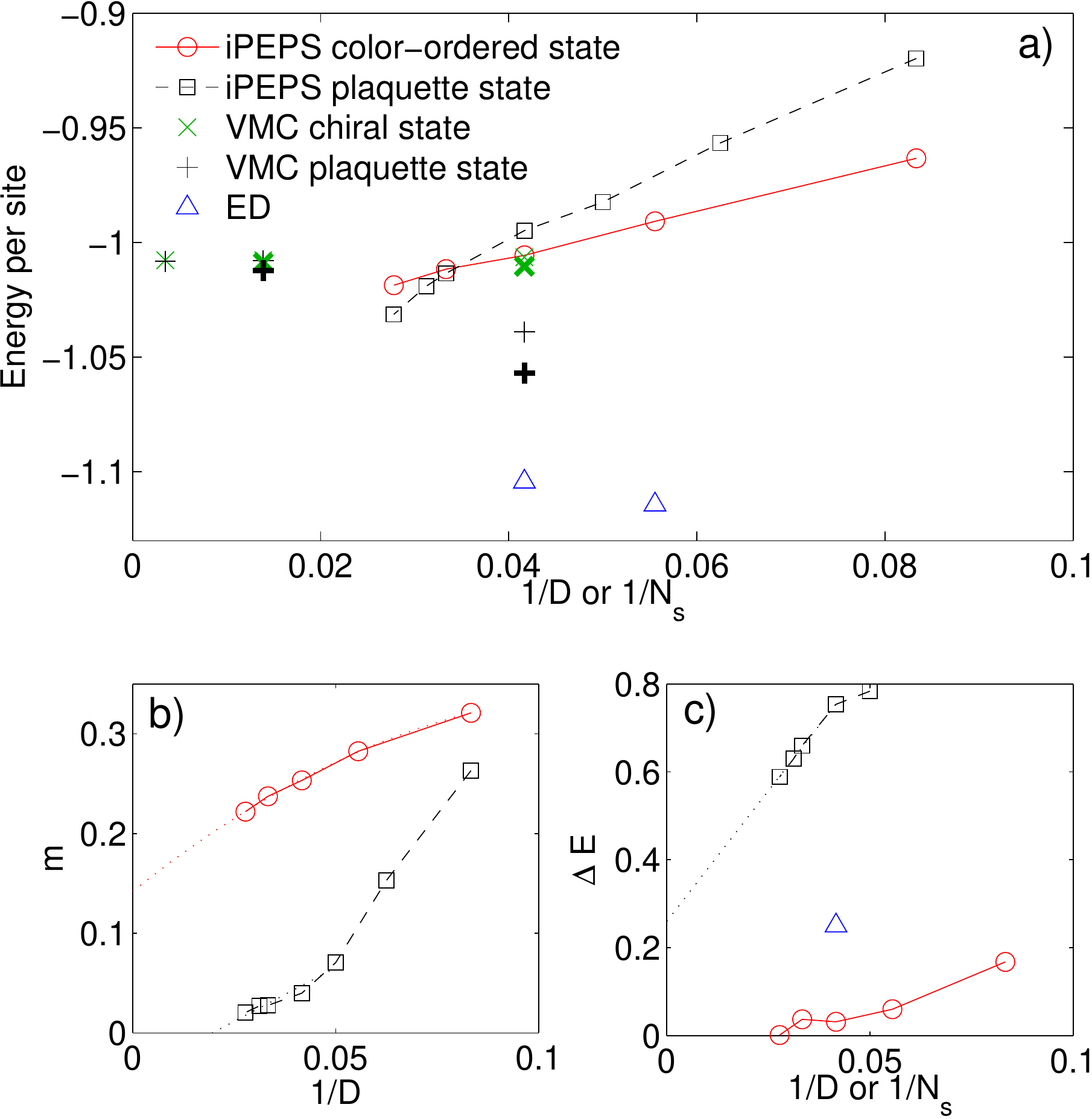}
\caption{iPEPS results for the SU(6) Heisenberg model on the honeycomb lattice. (a) Comparison of the ground state energy obtained with iPEPS, VMC, and ED, as a function of inverse bond dimension $D$ and inverse system size.  The bold symbols mark improved VMC results for $N_s=24$ and $72$ (see Table \ref{tab:VMCen} and main text). For large bond dimension with iPEPS the plaquette state has the lowest variational energy, in agreement with VMC.
(b) Color order parameter as a function of inverse~$D$. It is finite for the color-ordered state and  vanishes for the plaquette state. (c) Difference in energy between the strongest bond and the weakest bond in the unit cell, which is strongly suppressed in the color-order state, and finite in the plaquette state, consistent with plaquette long-range order. 
The dotted lines are a guide to the eye.
}
\label{PEPS}
\end{figure}

As an additional test, we have determined the spatial quantum numbers of the first excited doublet by applying one of the two elementary translations of the lattice. The corresponding eigenstates belong to the two $K$ points in the Brillouin zone. The correlations in these states are very similar to those in the ground state, which suggests that these three states could correspond to the degenerate ground state of the thermodynamic limit split by finite size effects. To demonstrate that this is the case, we have constructed the symmetric sum of these states, which corresponds to the finite-size approximation of a broken symmetry state (a simple task since the numerical wave functions are real and not complex). In that state, the strong bonds correspond to a covering of the lattice with hexagons (see inset of Fig.~\ref{ED_24sites}(b)), with a difference between strong and weak bond energies of $0.25$, in good agreement with the extrapolated iPEPS estimate (see below Fig. \ref{PEPS}(c)).

However, one should not forget that we have access to only one cluster with the appropriate number of low-lying states, and that the gap to the next levels is comparable to the gap between the ground state and the first pair of low lying states. So, in the next sections, we turn to the results obtained with other methods.

{\it iPEPS: }
An iPEPS is a variational tensor network ansatz to represent a 2D wavefunction in the thermodynamic limit.~\cite{verstraete2004,nishino01,jordan2008} The ansatz on the honeycomb lattice consists of a unit cell of rank-4 tensors which is periodically repeated on the infinite lattice, for each tensor one physical index carries the local Hilbert space of lattice site, and three auxiliary indices connect to the nearest-neighbor tensors. The accuracy of the ansatz can be systematically controlled by the bond dimension $D$ of the auxiliary indices. For the experts we note that the contraction of the tensor network is performed using a variant~\cite{corboz2011,corboz14_tJ} of the corner-transfer matrix method,~\cite{nishino1996, orus2009-1} and the optimization is done by an imaginary time evolution using a combined simple and (fast-) full update.~\cite{corboz2010, phien15} To increase the efficiency of the simulations we make use of abelian symmetries.~\cite{singh2010,bauer2011} A similar approach has been used in previous calculations of SU($N$) Heisenberg models, see e.g. Refs.~\onlinecite{Corboz13_su3hc,Corboz12_su4}. For an introduction to iPEPS we refer to Refs.~\onlinecite{corboz2010,phien15}.

We have used a 6-site unit cell which is compatible with both a plaquette state and a uniform state. As initial states we started either from completely random tensors or from a plaquette state made of SU(6) singlets on  hexagons. 
In the former case, using bond dimensions up to $D=24$, a new competing state appears, in which each site in the unit cell exhibits a different dominant color. 
For $D\leq 24$ this color ordered state has a lower variational energy than the plaquette state, as shown in Fig.~\ref{PEPS}(a). However, the slope in $1/D$ is larger for the plaquette state. So we have pushed the calculation to very large values of $D$, up to $D=36$. Around $D=30$ the energies of the two ordered states indeed cross such that the plaquette state clearly becomes energetically favored. We have not found a competing uniform chiral state with iPEPS which is an indication that at least for the bond dimensions studied here the plaquette state is the lowest energy state, consistent with the VMC result. 

In Fig.~\ref{PEPS}(b) we present the results for the color-order parameter of the two competing states, given by the local moment
\begin{equation}
\label{eq:m}
m=\sqrt{\frac{6}{5} \sum_{\alpha,\beta} \left(\langle S_\alpha^\beta \rangle - \frac{ \delta_{\alpha\beta}}{6 } \right)^2},
\end{equation}
averaged over all sites in the unit cell, where $S_\alpha^\beta = |\alpha\rangle\langle \beta|$ are the SU(6) spin operators and $\alpha,\beta$ run over all local basis states. For the color-ordered state $m$ is large for low $D$. It decreases with increasing $D$ but tends to a finite value in the infinite $D$ limit. The local moment of the plaquette state is much more strongly suppressed with increasing $D$, and vanishes in the large $D$ limit, consistent with a singlet without color order.

Figure~\ref{PEPS}(c) shows the difference between the highest and lowest bond energy in the unit cell which measures the magnitude of the plaquette order. For the color ordered state it is strongly suppressed with increasing $D$ and vanishes for large~$D$, in contrast to the plaquette state which exhibits a large difference in bond energy, where the strong bonds form hexagonal plaquettes.

\begin{figure}
\includegraphics[width=0.45\textwidth]{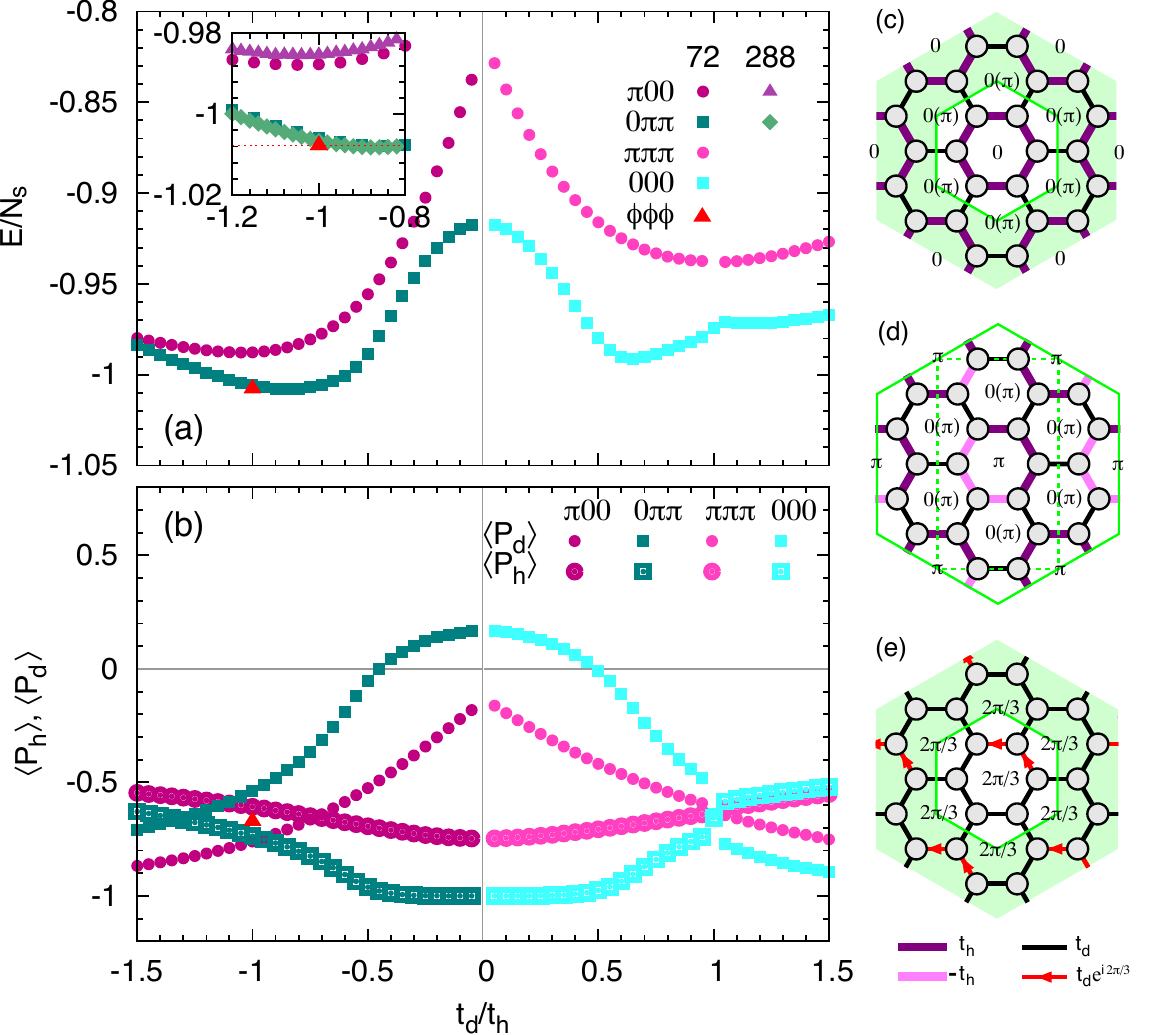}
\caption{ Energies of Gutzwiller projected wave functions (a) and the bond energies on $t_d$ and $t_h$ bonds after projection for the different flux configurations as a function of $t_d/t_h$ for $N_s=72$. (c)-(e) shows the considered flux configurations, black bonds represent  hopping amplitude $t_d$, while dark and light  purple bonds denote  hopping amplitudes $t_h$ and $-t_h$, respectively. In case of the uniform $2\pi/3$ flux configuration, red arrows represent complex hopping amplitude $\propto e^{i2\pi/3}$, for which $t_{ji}=t_{ij}^*$. 
}
\label{fig:td_vs_E}
\end{figure}

{\it VMC:}
Gutzwiller projected wave functions \cite{yokoyama1987,gros1989} offer a qualitative and potentially quantitative description for both types of competing scenarios found by mean-field study.\cite{SzirmaiG2011} 
In this method we project out the configurations having multiple occupancy from the Fermi-sea constructed from a mean-field model. The variational parameters are the hopping amplitudes and the artificial fluxes given by their total phase around the elementary hexagons (plaquettes). An importance sampling Monte Carlo method was used to calculate the energies and correlations of the projected states.\cite{Corboz12_su4}
Our calculations (shown in Fig.~\ref{fig:td_vs_E}) reveal that the lowest energy states are similar to those of Ref.~[\onlinecite{SzirmaiG2011}]:  (i) a configuration with uniform $2 \pi/3 $-flux before projection, corresponding to a chiral spin-liquid\cite{WenWiltzekPRB1989},
and (ii) a translation symmetry breaking configuration with $0$-flux in a center plaquette surrounded by $\pi$-flux plaquettes with non-uniform hopping integrals, corresponding to a plaquette ordered phase. 
While the mean-field results of Ref.~[\onlinecite{SzirmaiG2011}] slightly favored the chiral phase,
the plaquette-ordered phase turned out, after projection, to have a slightly lower energy for all studied system sizes (see Table \ref{tab:VMCen}), in agreement with the other numerical approaches.  

The energy minimum for the $0\pi\pi$-flux states, shown in Fig.~\ref{fig:td_vs_E}, occurs for $t_d/t_h\approx -0.85$. Now, for $t_d \leq -t_h/2$, which includes the optimal energy value, the fermionic wave function is gapless at the Fermi-energy: the lowest filled band touches the empty band above it at the $\Gamma$ point.\cite{Corboz13_su3hc}.  So, by contrast to the plaquette phase of the SU(3) Heisenberg on the honeycomb lattice, which is described by a gapped fermionic wave function\cite{Corboz13_su3hc}, the plaquette phase discussed here for SU(6) corresponds to a gapless spectrum before projection, hence possibly
also to a gapless spectrum after projection.  Since this gapless point is not protected (the spectrum is gapped for $t_d > -t_h/2$), we suspect that this is an artefact, and that adding additional terms in the fermionic Hamiltonian might open a gap and further lower the variational energy of that state, which is not
as good as that of the chiral state (see below). However, it might as well be that the spectrum is indeed gapless. This point deserves further investigation.

\begin{table}[htdp]
\begin{center}
\begin{tabular}{|c|c|c|c|c|c|c|}

\hline
$N_s$&24  & 24 opt & 72 & 72 opt & 288 & meanfield\cite{SzirmaiG2011}\\
\hline
plaquette &	-1.039& -1.057  & -1.0079&-1.0123& -1.0082	& -1.010  \\
\hline
$\frac{2\pi}{3}$ chiral &-1.0064 &-1.0104 &-1.0077& -1.0087 &	-1.0077& -1.025 \\ 
\hline
\end{tabular}
\end{center}
\caption{VMC energies of Gutzwiller projected wave functions for the competing $0\pi\pi$ (plaquette) and the $2\pi/3$ flux configurations for different system sizes. The statistical error of the calculations is smaller than $ \mathcal{O}(10^{-4})$.
The optimized energies are obtained by considering the overlap between projected states with different  boundary conditions before projection.
}
\label{tab:VMCen}
\end{table}%

{\it Ring exchange term:} 
Since the energy difference between the plaquette and chiral phases found by VMC is very small, it is tempting
to speculate that the chiral phase might be stabilized by a ring exchange term around the hexagons. We have thus considered
 \begin{equation}
\mathcal{H} =\cos \theta \sum_{\langle i,j \rangle} P_{ij} +\sin \theta \sum_{\mathrm{plaquttes}} i
\left (P_{\hexagon}^{\phantom{-1}} - P_{\hexagon}^{-1}\right)
\label{ring_exchange}
\end{equation}
where the sum in the second term runs over all hexagonal plaquettes, and the operators $P_{\hexagon}^{\phantom{-1}}$ and $P_{\hexagon}^{-1}$ permute the configuration on a hexagon clockwise and anticlockwise  (also called ring exchange terms). The new term directly couples to the scalar chirality on the hexagons, breaks time-reversal 
invariance, and is a bona-fide SU(6) generalization of an SU(2) Hamiltonian on the kagome lattice which has been shown to give rise to an extended SU(2) chiral spin liquid 
phase~\cite{Bauer2014,WietekLaeuchli2015}. Alternatively it can be viewed as a drastically truncated version of a parent Hamiltonian for a SU(N) chiral spin liquid~\cite{Tu2014328}.
 
In the following, we will discuss the properties of that model as a function of $\theta$, noting that $\theta=0$ corresponds to the pure Heisenberg model (\ref{eqn:HamP2}).

\begin{figure}[t]
\includegraphics[width=0.45\textwidth]{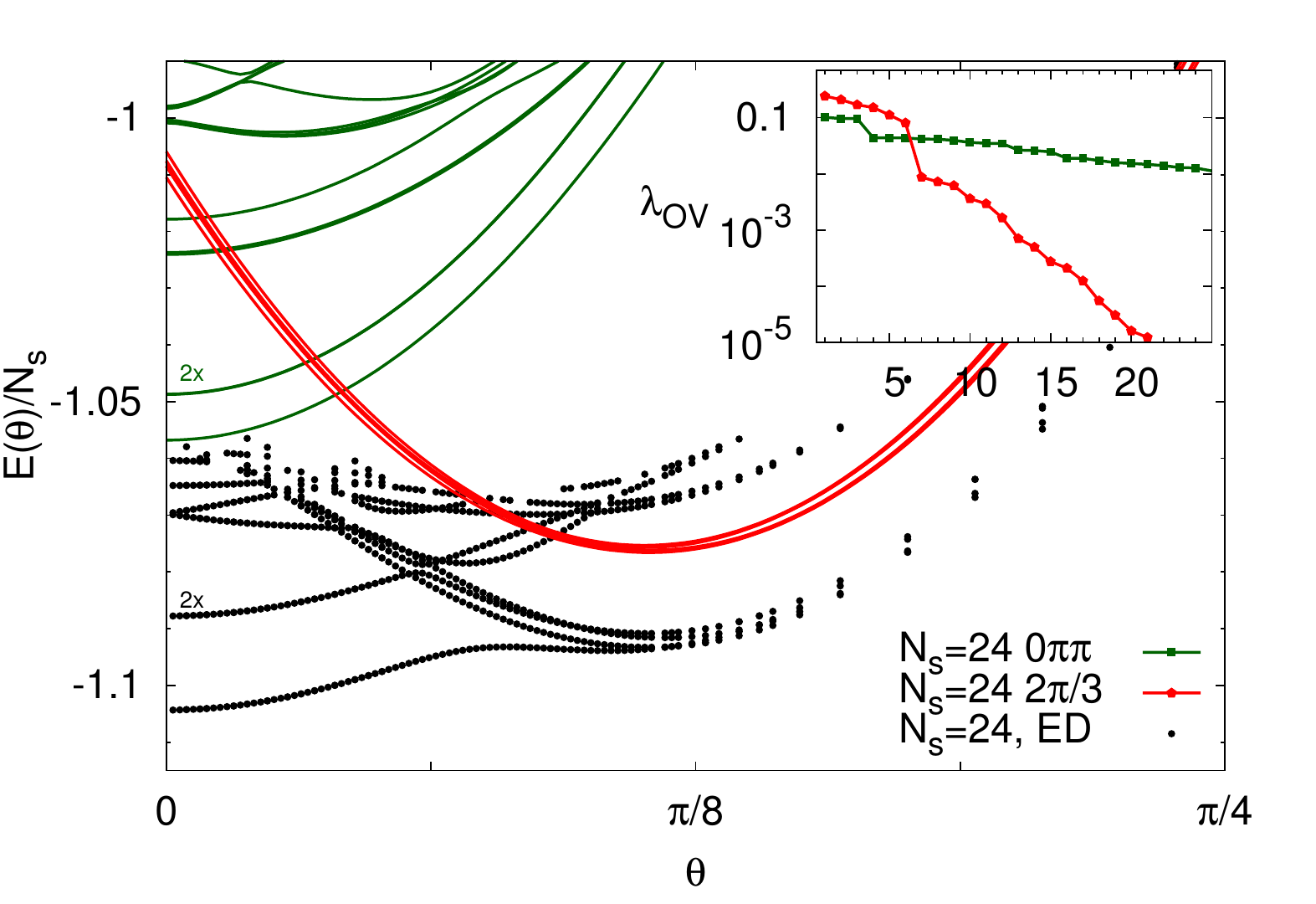}
\caption{Comparison of the ED spectrum (black points) of the model of Eq.~(\ref{ring_exchange}) with the variational energies (continuous lines) based on Gutzwiller projected wave-functions for the $0\pi\pi$ plaquette phase and the $2\pi/3$ chiral phase. The inset shows the  eigenvalues ( $\lambda_{\text{OV}}$ ) of the overlap  matrices of the projected states with different twisted boundary conditions before projection. 
}
\label{ED_VMC_Comparison}
\end{figure}

  The ED spectrum on 24 sites (Fig.~\ref{ED_VMC_Comparison}) shows a clear change of behavior  between the small $\theta$ range, with a twofold excited state 
 well separated from the rest of the spectrum, and the range above $\theta \simeq 0.2$, where a manifold of 6 singlet states becomes
 almost degenerate and very well separated from the rest of the spectrum. Two of these states are at the $\Gamma$ point, and the remaining four are at the $K$ points, in agreement with the momenta of the six chiral VMC states (discussed below). So, the ED results are clearly consistent with a phase transition between a plaquette phase and a chiral phase upon increasing the ring exchange term. Note that the degeneracy of the chiral
 state is only equal to 6 and not 12 because the Hamiltonian of Eq.~(\ref{ring_exchange}) explicitly breaks the time reversal symmetry. 
 
This interpretation is further supported by the comparison with VMC on 24 sites.  To access the low energy spectrum and not just the ground state, we have constructed a large family of Gutzwiller projected states by changing the boundary conditions (BC) of the fermionic wave-functions,\cite{ZhangOshikawaPRB2012} 
considering up to 30 different BCs for the $2\pi/3$ flux states, and up to 90 for the $0\pi\pi$-flux states (30 for each translation breaking state), and
we have diagonalized the overlap matrix and the Hamiltonian in this variational subspace.\cite{LiYangPRB2010,MeiWenarxiv2014} The results are summarized in Fig.~\ref{ED_VMC_Comparison}. For the chiral state,
this parton construction leads to 6 (and only 6) significant eigenvalues of the overlap matrix, which themselves lead to 6 low-lying states very close in energy\cite{Nataf_SUNchiral}, while for the plaquette states, there is not such a clear cutoff, and the three low-lying states are not so well split from the other states.
Although the variational plaquette and chiral states are higher in energy, their overall behavior is qualitatively consistent with ED. In particular, the energy of the plaquette state is minimal at $\theta = 0$, while that of the chiral states is minimal around $\theta=0.36$, and their energies cross around $\theta=0.16$. 

Similar overlap calculations were carried out for $N_s=72$ sites, with 30 different BCs for the $2\pi/3$ flux case, and 12 for each translation breaking state (36 in total) for the $0\pi\pi$-flux case. The energy corrections for the $0\pi\pi$ case turn out to be larger (see Table \ref{tab:VMCen}), again promoting the plaquette ordered phase over the chiral liquid phase at the Heisenberg point. \footnote{Note that while the diagonal energies depend on the value of $t_d/t_h$,  the spanned subspace of the projected states with different boundary conditions before projection remains the same, thus the optimized energies are independent of small changes of $t_d/t_h$.}

{\it Discussion: } To summarize, the numerical evidence clearly points to a plaquette ground state for the SU(6) model on the honeycomb lattice, but 
with a chiral phase close by in parameter space. Even if it led to the wrong conclusion, the mean-field approach should be given credit for identifying
the right candidates with very similar energies.\cite{SzirmaiG2011} This lends further support to the mean-field prediction by 
Hermele {\it et al.}\cite{hermele2009,HermeleGurarie2011} of a chiral phase for several particles per site since there does not seem to be competing VBS states too close in energy in that case.  Numerical work along the lines of the present paper to test this prediction is in progress.

{\it Acknowledgements:}  
This work has been supported by the Swiss National Science Foundation, the  JSPS KAKENHI Grant Number 2503802, by the Hungarian OTKA Grant No. K106047, by the National Science Foundation under Grant No. NSF PHY11-25915, by the Austrian Science Fund FWF (F-4018-N23 and I-1310-N27) and by the Delta-ITP consortium (a program of the Netherlands Organisation for Scientific Research (NWO) that is funded by the Dutch Ministry of Education, Culture and Science (OCW)). 

\bibliographystyle{apsrev4-1}
\bibliography{SU6_refs,refs_PC}

\end{document}